\documentclass[twocolumn,aps,showpacs,prb,tightenlines,amsmath,amssymb,superscriptaddress]{revtex4}
\usepackage{graphicx}
\usepackage{amssymb}
\usepackage{amsmath}
\usepackage{colordvi} 
\newcommand{\bgreek}[1]{\mbox{\boldmath$#1$\unboldmath}}
\begin{document} 

\title{Effect of spin-conserving scattering on Gilbert damping in ferromagnetic semiconductors}

\author{K.\ Shen}
\affiliation{Hefei National Laboratory for Physical Sciences at
  Microscale and Department of Physics, 
University of Science and Technology of China, Hefei,
  Anhui, 230026, China}
\author{G.\ Tatara}
\affiliation{Department of Physics, Tokyo Metropolitan University,
Hachioji, Tokyo 192-0397, Japan}
\author{M.\ W.\ Wu}
\thanks{Author to whom correspondence should be addressed}
\email{mwwu@ustc.edu.cn.}
\affiliation{Hefei National Laboratory for Physical Sciences at
  Microscale and Department of Physics, 
University of Science and Technology of China, Hefei,
  Anhui, 230026, China}
\date{\today}

\begin{abstract}
The Gilbert damping in ferromagnetic semiconductors is theoretically
investigated based on the $s$-$d$ model. 
In contrast to the situation in metals, all the spin-conserving
scattering in ferromagnetic semiconductors supplies an
additional spin relaxation channel due to the momentum dependent effective
magnetic field of the spin-orbit coupling,
thereby modifies the Gilbert damping.
In the presence of a pure spin current, we predict a new contribution
due to the interplay of the anisotropic spin-orbit coupling and a 
pure spin current. 
\end{abstract}


\maketitle

The ferromagnetic systems 
have attracted much attention both for the abundant
fundamental physics and promising applications in the past
decade.\cite{tserk-rev,tatara} The study on the collective
magnetization dynamics in such systems has been an active field with
the aim to control the magnetization. In the literature, the
magnetization dynamics is usually described by the phenomenological
Landau-Lifshitz-Gilbert (LLG) equation,\cite{llg}
\begin{eqnarray}
  {\dot {\bf n}}=\gamma{\bf H}_{\rm eff}\times{\bf n}+\alpha{\bf
    n}\times{\dot {\bf n}},
  \label{eq1}
\end{eqnarray}
with $\bf n$ denoting the direction of the magnetization. The first
and second terms on the right hand
side of the equation represent the precession and 
relaxation of the magnetization
under the effective magnetic field ${\bf H}_{\rm eff}$, respectively. The relaxation
term is conventionally named as the Gilbert damping term with the damping
coefficient $\alpha$. The time scale of the
magnetization relaxation then can be estimated by $1/(\alpha\gamma
H_{\rm eff})$,\cite{tserk1} which is an important parameter for dynamic manipulations. 
The coefficient $\alpha$ is essential in determining the efficiency of the current-induced magnetization swiching, and experimental determination of $\alpha$ has been carried out intensively in metals\cite{Oogane06}
and magnetic semiconductors.\cite{Adam09}

To date, many efforts have been made to clarify the microscopic origin
of the Gilbert damping.\cite{piechon,tserk2,kohno1,
  kohno2,kunes,steiauf} Kohno {\em et al.}
\cite{kohno1} employed the standard diagrammatic perturbation approach
to calculate the spin torque in the small-amplitude magnetization
dynamics and obtained a Gilbert torque with the damping coefficient
inversely proportional to 
the electron spin lifetime. They showed that the electron-non-magnetic
impurity scattering, a spin-conserving process, does not affect the Gilbert damping.
Later, they extended the theory into the finite-amplitude dynamics by
introducing an SU(2) gauge field\cite{tatara} and obtained a Gilbert
torque identical to that in the case of small-amplitude
dynamics.\cite{kohno2} In those calculations, the electron-phonon and
electron-electron scatterings were discarded.
One may infer that both of them should be irrelevant to the Gilbert
damping in ferromagnetic metals, since they are independent of 
the electron spin relaxation somewhat like the electron-non-magnetic impurity
scattering. 
However, the situation is quite different in
ferromagnetic semiconductors, where the spin-orbit coupling (SOC) due to the
bulk inversion asymmetry\cite{dress} and/or the structure inversion
asymmetry\cite{rashba} presents
a momentum-dependent effective magnetic field (inhomogeneous
broadening\cite{wu1}). As a result, any spin-conserving scattering, including
the electron-electron Coulomb scattering, can
result in a spin relaxation channel to affect the Gilbert
damping. In this case, many-body effects on the Gilbert damping due to
the electron-electron Coulomb scattering should be expected.
Sinova {\em et al.}\cite{sinova} studied the Gilbert damping in GaMnAs
ferromagnetic semiconductors by including the SOC 
to the energy band structure. In that work,
the dynamics of the carrier spin coherence was missed.\cite{tserk3}
The issue of the present work is to study the Gilbert
damping in a coherent frame.

In this Report, we apply the gauge field approach to
investigate the Gilbert damping
in ferromagnetic semiconductors. In our frame, all the relevant
scattering processes, even the electron-electron scattering which
gives rise to many-body effects, can be included. The goal of this work
is to illustrate the role of the SOC and
spin-conserving scattering on Gilbert damping. We
show that the spin-conserving scattering can affect
the Gilbert damping due to the contribution on spin relaxation process. We
also discuss the case with a pure spin current, from which we predict
a new Gilbert torque due to the 
interplay of the SOC and the spin current.

Our calculation is based on the $s$-$d$ model with itinerant $s$ and
localized $d$ electrons. The collective magnetization arising from the $d$
electrons is denoted by ${\bf M}=M_s{\bf n}$. The exchange
interaction between itinerant and localized electrons can be written as
$H_{\rm sd}=M\int d{\bf r} ({\bf n}\cdot {\bgreek \sigma})$,
where the Pauli matrices ${\bgreek \sigma}$ are spin operators of the
itinerant electrons and $M$ is the coupling constant. In order to
treat the magnetization dynamics with an arbitrary amplitude,\cite{kohno2}
we define the temporal spinor operators of the itinerant electrons
$a(t)=(a_\uparrow(t),a_\downarrow(t))^T$ in the rotation coordinate system with 
$\uparrow$ ($\downarrow$) labeling the spin orientation parallel
(antiparallel) to $\bf n$. With a unitary transformation
matrix $U(t)$, one can connect the operators $a_{\uparrow(\downarrow)}$ with those
 defined in the lattice 
coordinate system $c_{\uparrow(\downarrow)}$ by $a(t)=U(t)c$. 
Then, an SU(2) gauge field $A_\mu(t)=-iU(t)^\dag
(\partial_\mu U(t))={\bf A}_\mu(t)\cdot{\bgreek 
  \sigma}$ should be introduced into the rotation framework to
guarantee the invariance of the total Lagrangian.\cite{kohno2} In the
slow and smooth precession limit, the gauge field can 
be treated perturbatively.\cite{kohno2} Besides, one
needs a time-dependent $3\times 3$ orthogonal rotation matrix ${\cal R}(t)$, which
obeys $U^\dag {\bgreek \sigma}U={\cal R} {\bgreek \sigma}$, to
transform any vector between the two coordinate systems. More details
can be found in Ref.\ \onlinecite{tatara}. In the following, we
restrict our derivation to a spatially homogeneous system, 
to obtain the Gilbert damping torque.

Up to the first order, the interaction Hamiltonian due to the gauge
field is  $H_A=\sum_{{\bf k}} {\bf A}_{0}\cdot a_{{\bf
      k}}^\dag{\bgreek \sigma} a_{\bf k}$
and the spin-orbit couping reads
\begin{equation}
  H_{so}=\tfrac{1}{2}\sum_{\bf k}{\bf h}_{\bf k}\cdot c^\dag{\bgreek
    \sigma}c=\tfrac{1}{2}\sum_{\bf k}\tilde{{\bf
      h}}_{\bf k}\cdot a_{\bf k}^\dag{\bgreek
    \sigma}a_{\bf k},
\label{eq2}
\end{equation}
with $\tilde{\bf h}={\cal R}{\bf h}$. Here, we
take the Planck constant $\hbar=1$. 
We start from the fully microscopic kinetic spin Bloch equations
 of the itinerant electrons derived from the non-equilibrium 
Green's function approach,\cite{wu1,wu2}
\begin{equation}
  \partial_t\rho_{\bf k}=\partial_t\rho_{\bf k}\big|_{\rm
    coh}+\partial_t\rho_{\bf k}\big|_{\rm
    scat}^{c}+\partial_t\rho_{\bf k}\big|_{\rm scat}^{f},
  \label{eq3}
\end{equation}
where $\rho_{\bf k}$ represent the itinerant electron density matrices
defined in the rotation coordinate system. The
coherent term can be written as 
\begin{equation}
  \partial_t\rho_{\bf k}\big|_{\rm coh}=-i[{\bf \cal A}\cdot {\bgreek
    \sigma},\rho_{\bf k}]-i[\tfrac{1}{2}{\tilde{\bf h}_{\bf k}}\cdot{\bgreek 
    \sigma}+\hat\Sigma_{\rm HF},\rho_{\bf k}].
  \label{eq4}
\end{equation}
Here $[ , ]$ is the commutator and ${\bf \cal A}(t)={\bf A}_0(t)+M\hat
{\bf z}$ with ${\bf A}_0$ and $M\hat{\bf z}$ representing the gauge field and
effective magnetic filed due to $s$-$d$ exchange interaction, respectively.
$\hat\Sigma_{\rm HF}$ is the
Coulomb Hartree-Fock term of the 
electron-electron interaction.
$\partial_t\rho_{\bf k}\big|_{\rm scat}^{c}$ and $\partial_t\rho_{\bf
  k}\big|_{\rm scat}^{f}$ in Eq.\,(\ref{eq3}) include all the
relevant spin-conserving and spin-flip scattering processes, respectively. 

The spin-flip term $\partial_t\rho_{\bf  k}\big|_{\rm scat}^{f}$
results in the damping effect was studied in Ref. \onlinecite{kohno2}.
Let us confirm this by considering the case of the 
magnetic disorder $V_{\rm imp}^m =u_s\sum_j\tilde{\bf S}_j\cdot a^\dag{\bgreek
  \sigma}a\delta({\bf r}-{\bf R}_j)$.
The spin-flip part then reads
\begin{equation}
  \partial_t\rho_{\bf k}\big|_{\rm scat}^f=\partial_t\rho_{\bf
    k}\big|_{\rm scat}^{f(0)}+\partial_t\rho_{\bf
    k}\big|_{\rm scat}^{f(1)},
  \label{eq5}
\end{equation}
with $\partial_t\rho_{\bf k}\big|_{\rm scat}^{f(i)}$ standing for the
$i$-th order term with respect to the gauge field, i.e.,
\begin{widetext}
\begin{eqnarray}
  \partial_t\rho_{\bf k}\big|_{\rm scat}^{f(0)}&=&-\frac{\pi n_su_s^2S_{\rm imp}^2}{3}
    \sum_{{\bf k}_1\eta_1\eta_2}\sigma^\alpha
    \rho_{{\bf k}_1}^>(t)T_{\eta_1}\sigma^\alpha T_{\eta_2}\rho_{\bf
      k}^<(t)\delta(\epsilon_{{\bf k}_1\eta_1}-\epsilon_{{\bf
        k}\eta_2})-(>\leftrightarrow <)+{\rm H.c.},\\
  \partial_t\rho_{\bf k}\big|_{\rm
    scat}^{f(1)}&=&\frac{i2\pi n_su_s^2S_{\rm imp}^2}{3}
  \varepsilon^{\alpha\beta\gamma}A_0^\gamma(t)\sum_{{\bf
      k}_1\eta_1\eta_2}\sigma^\alpha 
  \rho_{{\bf k}_1}^>(t)T_{\eta_1}\sigma^\beta T_{\eta_2}\rho_{\bf
    k}^<(t)\frac{d}{d\epsilon_{{\bf
        k}_1\eta_1}}\delta(\epsilon_{{\bf k}_1\eta_1}-\epsilon_{{\bf
        k}\eta_2})-(>\leftrightarrow <)+{\rm H.c.},
  \label{eq6-7}
\end{eqnarray}
\end{widetext}
where $T_\eta(i,j)=\delta^{\eta i}\delta^{\eta j}$ for the spin band
$\eta$. Here $\rho_{\bf k}^>=1-\rho_{\bf k}$, $\rho_{\bf
  k}^<=\rho_{\bf k}$.  $(>\leftrightarrow <)$ is obtained by 
interchanging $>$ and $<$
from the first term in each equation. $\varepsilon^{ijk}$ is the Levi-Civita
permutation symbol. The gauge field term,
$\partial_t\rho_{\bf k}\big|_{\rm  scat}^{f(1)}$, results from the spin
correlation of a single magnetic impurity at different
times.\cite{kohno2}
It induces a spin polarization proportional to 
$\hat{\bf z}\times{\bf A}_0^\perp(t)$ which gives a Gilbert torque. 
The damping coefficient is inversely proportional to the spin
relaxation time $\tau_{s}$ detemined by the spin-flip scattering 
$\partial_t\rho_{\bf
  k}\big|_{\rm scat}^{f(0)}$. The spin-flip scattering term in
Eq.\,(\ref{eq3}) thus reproduces the result of Ref. \onlinecite{kohno2}.

We now demonstrate that  the Gilbert damping torque arises also from 
the spin-conserving scattering.
For the discussion of the spin-conserving term, 
it is sufficient to  approximate the spin-flip term as
$\partial_t\rho_{\bf
  k}\big|_{\rm scat}^{f}=-{(\rho_{\bf k}-\rho^e_{\bf
    k})}/{\tau_s}$, with $\rho^e_{\bf k}$ representing the
instantaneous equilibrium distribution
(i.e., $\rho^e_{\bf k}$ is $\rho_{\bf k}$ without the gauge field and ${\bf P}^s_{\bf k}$). 
Equation\,(\ref{eq3}) then reads
\begin{eqnarray}
  \nonumber
  \partial_t\rho_{\bf k}&=&-i[{\cal A}\cdot{\bgreek \sigma},\rho_{\bf
    k}]-i[\tfrac{1}{2}\tilde{\bf h}_{\bf k}\cdot{\bgreek
    \sigma},\rho_{\bf k}] \\
  &&\mbox{}+\partial_t\rho_{\bf k}\big|_{\rm scat}^{c}-{(\rho_{\bf k}-\rho^e_{\bf
      k})}/{\tau_s}+{\bf P}^s_{\bf k}. 
  \label{eq8}
\end{eqnarray}
Here, we add an additional term, ${\bf P}^s_{\bf k}$, to describe the source
of a pure spin current due to the magnetization dynamic
pumping\cite{tserk1} or electrically injection\cite{jedema,lou} in
order to discuss the system with a pure spin current.
We neglect the Coulomb Hartree-Fock effective magnetic field
since it is approximately parallel to the $s$-$d$ exchange
field, but with a smaller magnitude.

By averaging density matrices over the momentum direction, one obtains
the isotropic component $\rho_{i, k}=\int \frac{d\Omega_{\bf
    k}}{4\pi}\rho_{\bf k}$. The
anisotropic component is then expressed as $\rho_{a,\bf k}=\rho_{\bf k}-\rho_{i,k}$.
It is obvious that this anisotropic component does not give any spin
torque in the absence of the SOC, since $\sum_{\bf k}{\rm 
  Tr}({\bgreek \sigma}\rho_{a,\bf k})=0$. 
Below, it is shown that this component leads to the damping when coupled to the spin-orbit interaction.

By denoting the isotropic component of the equilibrium part 
($\rho_{\bf k}^e$) as $\rho_{i,k}^e$ and representing the 
non-equilibrium isotropic part by
 $\delta\rho_{i,k}=\rho_{i,k}-\rho_{i,k}^e$, 
we write the kinetic spin Bloch equations of the non-equilibrium
isotropic density matrices
$\delta\rho_{i,k}$ 
and those of the anisotropic components $\rho_{a,\bf k}$ as
\begin{eqnarray}
  \nonumber
  \partial_t\rho_{i,k}&=&-\tfrac{\delta\rho_{i,k}}{\tau_s}-i[{\cal A}\cdot {\bgreek
    \sigma},\delta\rho_{i,k}]-i\overline{[\tfrac{1}{2}\tilde{\bf
      h}_{\bf k}\cdot{\bgreek \sigma},\rho_{a,\bf
      k}]}  \\
  &&\mbox{}-i[{\bf A}_0\cdot {\bgreek
    \sigma},\rho_{i,k}^e],\\
  \nonumber
  \partial_t\rho_{a,\bf k}&=&\partial_t\rho_{a,\bf 
      k}\big|_{\rm scat}^{c}-i[{\cal A}\cdot{\bgreek \sigma},\rho_{a,\bf
    k}]-i[\tfrac{1}{2}\tilde{\bf h}_{\bf k}\cdot{\bgreek
    \sigma},\delta\rho_{i,k}]\\
  &&\mbox{}-i[\tfrac{1}{2}{\tilde{\bf h}_{\bf
      k}}\cdot{\bgreek \sigma},\rho_{a,\bf
    k}]+i\overline{[\tfrac{1}{2}\tilde{\bf h}_{\bf k}\cdot{\bgreek 
      \sigma},\rho_{a,\bf k}]} +{\bf P}^s_{\bf k},
  \label{eq9-10}
\end{eqnarray}
respectively. The overline in these equations
presents a angular average over the momentum space. 

We further define  $\rho^{(0)}_{a,\bf k}$ as the anisotropic density 
in the absence of the gauge field, ${\bf A}_0$. 
As easily seen, it vanishes when ${\bf P}^s_{\bf k}=0$. 
The anisotropic component involving the
gauge field is denoted by  
$\rho_{a,\bf
  k}^{(1)} =\rho_{a,\bf k}-\rho^{(0)}_{a,\bf k}$. 
Equation (\ref{eq9-10}) is expressed in terms of these components as
\begin{eqnarray}
  \nonumber
  \partial_t\rho^{(0)}_{a,\bf k}&=&-i[{\bf M}\cdot{\bgreek \sigma},\rho^{(0)}_{a,\bf
    k}] +\partial_t\rho^{(0)}_{a,\bf 
      k}\big|_{\rm scat}^{c}+{\bf  P}^s_{\bf k}\\
    &&\mbox{}-i[\tfrac{1}{2}{\tilde{\bf h}_{\bf
      k}}\cdot{\bgreek \sigma},\rho^{(0)}_{a,\bf
    k}]+i\overline{[\tfrac{1}{2}\tilde{\bf h}_{\bf k}\cdot{\bgreek
      \sigma},\rho^{(0)}_{a,\bf k}]},\\
  \nonumber
  \partial_t\rho_{a,\bf k}^{(1)}&=&\partial_t\rho_{a,\bf 
      k}^{(1)}\big|_{\rm scat}^{c}-i[{\cal A}\cdot{\bgreek \sigma},\rho_{a,\bf
    k}^{(1)}]-i[\tfrac{1}{2}\tilde{{\bf h}}_{\bf k}\cdot{\bgreek
    \sigma},\delta\rho_{i,k}]\\
  &&  \mbox{}-i[\tfrac{1}{2}\tilde{{\bf h}}_{\bf k}\cdot{\bgreek
    \sigma},\rho_{a,\bf k}^{(1)}]-i[{\bf A}_0\cdot{\bgreek \sigma},\rho^{(0)}_{a,\bf
    k}].
  \label{eq11-12}
\end{eqnarray}
Within the elastic scattering approximation, the electron-phonon
scattering as well as the
electron-non-magnetic impurity scattering can be simply written as
$\sum_{l,m}\rho_{a,k,lm}^{(1)}Y_{lm}/\tau_l$, where the
density matrices are expanded by the spherical harmonics functions
$Y_{lm}$, i.e., $\rho_{a,k,lm}^{(1)}=\int\frac{d\Omega_{\bf k}}{4\pi} \rho_{a,\bf
  k}^{(1)}Y_{lm}$. $\tau_l$ is the effective momentum relaxation time. 
The exact calculation of the Coulomb scattering is more complicated.
Nevertheless,  one can still express this term in the form of $\rho_{a,\bf
  k}^{(1)}/F_{\bf k}(\rho)$, where $F_{\bf k}$ is a function of the
density matrices\cite{glazov} and reflects many-body effects. For simplification, we
just introduce a uniform momentum relaxation time $\tau_l^\ast$ in the
following calculation.
Expanding Eq.\,(\ref{eq11-12}) by the spherical harmonics functions, one obtains
\begin{eqnarray}
  \nonumber
  &&\hspace{-0.5cm}\partial_t\rho_{a,k,lm}^{(1)}=-i[{\cal A}\cdot{\bgreek \sigma},\rho_{a,k,
    lm}^{(1)}]-i[\tfrac{1}{2}\tilde{{\bf h}}_{k,lm}\cdot{\bgreek
    \sigma},\delta\rho_{i,k}]\\
    &&\mbox{}-i[{\bf A}_0\cdot{\bgreek \sigma},\rho^{(0)}_{a,k,lm}]
  -i[\tfrac{1}{2}\tilde{{\bf h}}_{\bf k}\cdot{\bgreek 
    \sigma},\rho_{a,\bf k}^{(1)}]_{lm}
  -\tfrac{\rho_{a,k,lm}^{(1)}}{\tau_l^\ast},
  \label{eq13}
\end{eqnarray}
where the expanssion coefficient of any term $f_{\bf k}$ is expressed as
$f_{k,lm}=\int\frac{d\Omega_{\bf k}}{4\pi}f_{\bf k}Y_{lm}$.
In the strong scattering regime, i.e., $\frac{1}{\tau_l^\ast}\gg M $
and $\frac{1}{\tau_l^\ast}\gg |{\bf h}_{\bf k}|$, the
first and fourth terms are much smaller than the last term, hence can 
be discarded from the 
right side.  By taking the fact that the time derivative is a 
higher order term into account,
one also neglects $\partial_t\rho_{a,k,lm}^{(1)}$.
The solution of Eq.\,(\ref{eq13}) can be written as
\begin{equation}
  \rho_{a,k,lm}^{(1)}=-i\tau_l^\ast\{
    [\tfrac{1}{2}\tilde{\bf h}_{k,lm}\cdot{\bgreek \sigma},\delta\rho_{i,
      k}]+[{\bf A}_0\cdot{\bgreek \sigma},\rho^{(0)}_{a,k,lm}]\}.
    \label{eq14}
\end{equation}
Substituting it into Eq.\,(\ref{eq14}) and rewriting the equation in
the leading order, one obtains
\begin{eqnarray}
  \nonumber
&&\hspace{-0.2cm} \partial_t\rho_{i,k}=-i[{\cal A}\cdot {\bgreek
    \sigma},\delta\rho_{i, k}]-\tfrac{i}{2}\overline{[\tilde{\bf
      h}_{\bf k}\cdot{\bgreek \sigma},\rho^{(0)}_{a,\bf
      k}]}-i[{\bf A}_0\cdot {\bgreek
    \sigma},\rho_{i,k}^e]\\
&& \mbox{}-\sum_{lm}\tfrac{\tau_l^\ast}{4}{\left[\tilde{\bf
      h}_{k,lm}\cdot{\bgreek \sigma},
    [\tilde{\bf h}_{k,lm}\cdot{\bgreek \sigma},\delta\rho_{i,
      k}]\right]}-\tfrac{\delta\rho_{i,k}}{\tau_s}.
\label{eq15}
\end{eqnarray}
The third term on the right hand
side of the equation is proportional to the second order term of the 
SOC, which gives the spin dephasing channel  
due to the D'yakonov-Perel' (DP) mechanism.\cite{dp} This term can be
expressed by $\tau_{DP}^{-1}\delta\rho_{i,k}$ with
$\tau_{DP}^{-1}$ standing for the spin dephasing rate
tensor, which can be written as 
  $(\tau_{DP}^{-1})_{i,j}=\sum_{l,m}\langle\tau_l^\ast({({\bf
      h}_{k,lm})^2}\delta_{ij}-{h_{k,lm}^i h_{k,lm}^j})\rangle$
by performing the ensemble
averaging over the electron distribution. In the following, we treat
$\tau_{DP}$ as a scalar for simplification and 
the total spin lifetime is hence given by
\begin{eqnarray}
\tau_r=1/(\tau_{DP}^{-1}+\tau_s^{-1}).
\label{eq16}
\end{eqnarray}
Similar to the previous procedure, we
discard $\partial_t\rho_{i,k}$ in Eq.\ (\ref{eq15}) and obtain
\begin{equation}
 i[{\cal A}\cdot {\bgreek
    \sigma},\delta\rho_{i, k}]+{\delta\rho_{i,k}}/{\tau_r}
=-i\overline{[\tfrac{1}{2}\tilde{\bf
      h}_{\bf k}\cdot{\bgreek \sigma},\rho^{(0)}_{a,\bf
      k}]}-i[{\bf A}_0\cdot {\bgreek
    \sigma},\rho_{i,k}^e].
  \label{eq17}
\end{equation}
By taking $\delta\tilde{\bf s}_i=\frac{1}{2}\sum_{\bf k}{\rm
  Tr}({\bgreek \sigma}\delta\rho_{i,k})$, $\tilde{\bf
  s}_i^e=\frac{1}{2}\sum_{\bf k}{\rm Tr}({\bgreek \sigma}\rho^e_{i, 
  k})$ and $\tilde{\bf s}_{a,\bf
  k}^{(0)}=\frac{1}{2}{\rm Tr}({\bgreek \sigma}\rho^{(0)}_{a,\bf
  k})$, one can write the solution as
\begin{equation}
  \delta\tilde{\bf s}_i=\frac{\tilde{\bf v}+2\tau_r{\cal
      A}\times\tilde{\bf v}+ 4\tau_r^2\left(
      \tilde{\bf v}\cdot {\cal A}
    \right){\cal A} }{1+4|{\cal
      A}|^2\tau_r^2}-\tilde{\bf s}_i^e,
\label{eq18}
\end{equation}
where  $\tilde{\bf v}=\tilde{\bf s}_i^e+\tau_r\sum_{\bf k}{\tilde{\bf
    h}_{\bf  k}\times \tilde{\bf s}^{(0)}_{a,\bf k}}$.
$\tilde{\bf s}_i^e$ is just the equilibrium spin density
, which is parallel to the magnetization,
i.e., $\tilde{\bf s}_i^e=\tilde{s}_i^e\hat{\bf z}$. Now, we pick up
the transverse component in the form of $\hat {\bf z }
\times {\bf A}_0^\perp$, $\delta\tilde{\bf s}^\perp$, since only this
component results in a Gilbert torque of the magnetization as
mentioned above. We come to 
\begin{equation}
  \delta\tilde{\bf s}^\perp=2\tilde v_z({\bf
    A}_0^\perp\times\hat{\bf  z})\tau_{\rm ex}^2\tau_r/(\tau_r^2+\tau_{\rm ex}^2),
\label{eq19}
\end{equation}
with $\tau_{\rm ex}=1/(2M)$. By transforming it back to the lattice
coordinate system with ${\cal R}({\hat{\bf z}\times{\bf
    A}_0^\perp})=\tfrac{1}{2}\partial_t{\bf n}$,\cite{kohno2} one
obtains
\begin{equation}
  \delta{\bf s}^\perp=-\tilde v_z(\partial_t{\bf n})\tau_{\rm ex}^2\tau_r/({\tau_r^2+\tau_{\rm
      ex}^2}),
\label{eq20}
\end{equation}
This nonequilibrium spin polarization results in a spin torque
performed on the magnetization according to ${\bf
  T}=-2M{\bf n}\times \delta{\bf s}$, i.e.,
\begin{equation}
  {\bf T}=\tilde v_z({\bf n}\times\partial_t {\bf n})\tau_{\rm ex}\tau_r/(\tau_r^2+\tau_{\rm
      ex}^2).
  \label{eq21}
\end{equation}
Compared with Eq.\,(\ref{eq1}), the modification of the Gilbert
damping coefficient from this torque is
\begin{equation}
\alpha= \tilde v_z\tau_{\rm ex}\tau_r/(M_s\tau_r^2+M_s\tau_{\rm
    ex}^2),
\label{eq22}
\end{equation}

We fisrt discuss the case without the source term of the 
spin current.  In this case, the anisotropic
component $\rho_{a,\bf k}^{(0)}$ vanishes and $\tilde
v_z=\tilde s_{i}^e$. 
We see that the Gilbert damping then arises from $1/\tau_r$ [Eq.~(\ref{eq16})],
 i.e., from both the spin-flip scattering and the DP mechanism.\cite{dp}
Our main message is that this DP contribution is affected by the 
spin-conserving scattering processes such as the electron-electron 
interaction and phonons.
The temperature dependence of the Gilbert damping and the current-induced magnetization switching can thus be discussed quantitatively by evaluating $\tau_r$.
We note that our result reduces to the results of previous
works\cite{kohno2,piechon} when only the spin-flip scattering is considered.

We should point out that our formalism applies also to metals, 
by considering the case  $\frac{1}{\tau_l^\ast}\ll M $.
In this case, the last term of Eq.\,(\ref{eq13}) can be neglected and 
the effect of the spin-conserving scattering through $\tau_l^\ast$ becomes 
irrelevant.

When the pure spin current is included, we found additional 
contribution due to the interplay of the spin current and the SOC,
 since we have
\begin{equation}
  \tilde v_z=\tilde{s}_{i}^e+\Big[{\cal R}\Big(\tau_r\sum_{\bf k}{{\bf
          h}_{\bf  k}\times {\bf s}^{(0)}_{a,\bf
          k}}\Big)\Big]_z=\tilde{s}_{i}^e+\tilde s_z^{\rm sc},
  \label{eq23}
\end{equation}
with the spin current associated term $\tilde s_z^{\rm sc}$ defined
accordingly. The origin of $\tilde s_z^{\rm sc}$ can be understood
as follows. The anisotropic spin polarization ${\bf
  s}^{(0)}_{a,\bf k}$ arising from the pure spin current rotates
around the SOC effective magnetic field 
${\bf h}_{\bf k}$, which is also anisotropic. This precession finally results in
an isotropic spin polarization ${\bf s}^{\rm sc}=\tau_r\sum_{\bf k}{{\bf
    h}_{\bf  k}\times {\bf s}^{(0)}_{a,\bf k}}$ in the presence of spin
relaxation. This term contributes to the spin
polarization of the itinerant electrons along the direction of the
magnetization, i.e., $\tilde s_z^{\rm sc}$, thereby modifies the
Gilbert damping term by $\tilde s_z^{\rm sc}/\tilde s_{i}^e$.

The additional Gilbert damping due to the spin current found here is
different from the enhancement of the damping in the spin pumping
systems, where the existence of the interface is
essential.\cite{tserk1} 
In other words, what contributes there is the divergence of the spin current, as is understood from the continuity equation for the spin, indicating that the spin damping is equal to 
$\nabla\cdot {\bf j}_{s}+\dot{s}$ ($s$ is the total spin density).
In contrast, the damping found in the present paper arises even when the spin current is uniform if the spin-orbit interaction is there.

In summary, we have shown that the
spin-conserving scatterings in ferromagnetic
semiconductors, such as the
electron-electron, electron-phonon and electron-non-magnetic 
impurity scatterings,
contribute to the Gilbert damping in the presence of
the SOC because of the inhomogeneous broadening effect. 
We also predict that a Gilbert torque
arises from a pure spin current when coupled to the spin-orbit interaction.

This work was supported by the Natural Science Foundation of China
under Grant No.~10725417, the
National Basic Research Program of China under Grant 
No.~2006CB922005, the Knowledge Innovation Project of Chinese Academy
of Sciences,
Kakenhi (1948027) MEXT Japan and the 
Hitachi Sci. Tech. Foundation.

\end{document}